\documentclass[%
reprint,
superscriptaddress,
 amsmath,amssymb,
 aps,
]{revtex4-2}

\pdfoutput=1
\usepackage{graphicx}
\usepackage{dcolumn}
\usepackage{bm}
\usepackage{amsmath}
\usepackage[utf8]{inputenc}
\usepackage[T1]{fontenc}
\usepackage{mathptmx}
\usepackage{textcomp}
\usepackage{gensymb}
\usepackage{tensor}
\usepackage{xcolor}
\usepackage{braket}
\usepackage{caption}
\usepackage[normalem]{ulem}

\usepackage{empheq,etoolbox}
\patchcmd{\subequations}
  {\theparentequation\alph{equation}}
  {\theparentequation.\alph{equation}}
  {}{}
\begin{document}

\preprint{APS/123-QED}
 
\title{Micromotion-Synchronized Pulsed Doppler Cooling of Trapped Ions}

\affiliation{University of Washington, Department of Physics, Seattle, Washington, USA, 98195}

\affiliation{Physics Department, Brookhaven National Laboratory, Upton, New York, USA, 11973}

\author{Alexander Kato}
\affiliation{University of Washington, Department of Physics, Seattle, Washington, USA, 98195}

\author{Andrei Nomerotski}
\affiliation{Physics Department, Brookhaven National Laboratory, Upton, New York, USA, 11973}

\author{Boris B. Blinov}
\affiliation{University of Washington, Department of Physics, Seattle, Washington, USA, 98195}
\date{\today}

\begin{abstract}
 We propose and demonstrate a new method for Doppler cooling trapped-ion crystals where the distribution of micromotion amplitudes may be large and uneven. The technique uses pulses of Doppler cooling light synchronized with the trap RF that selectively target ions when their velocity is near a node, leading to more uniform cooling across a crystal by a single tone of cooling light. We lay out a theoretical framework that describes where this technique is practical, and provide a simple experimental demonstration.

\end{abstract}

\maketitle

\section{Introduction}

Micromotion is a time-dependent, driven motion that is present in all radiofrequency (RF) ion traps. This  may push ions beyond the Lamb-Dicke regime~\cite{Liebfried2003,Berkeland1998}, leading to frequency and amplitude modulation in addressing beams~\cite{Wang2015} and presenting a significant obstacle to high-fidelity quantum logic operations. Significant micromotion may even cause difficulties in Doppler cooling--typically a straightforward process for trapped ion crystals~\cite{Berkeland1998,Devoe1989}. For linear ion traps, this RF-driven motion can be minimized~\cite{Raizen1992,Prestage1989} for all ions by placing them at the nodal line where the RF electric field amplitude is zero, making this type of traps the most widely used by the experimenters.

However, several  recent proposals suggest 2D and 3D crystals may be used for quantum computing and simulations~\cite{Wang2015,Wang2020,Yoshimura2015,Richerme2016,Wu2020}. These proposals allude to the possibility of scaling up the number of qubits in a given area, opening up the potential for a wide range of quantum simulations that are more suited to a native 2D geometry and leading to new options for error correction that may improve the threshold for fault-tolerant quantum computing. To overcome the adverse effect of micromotion on gate fidelity, it has been shown that segmented pulses may be used~\cite{Wang2015,Wang2020}. Moreover, quantum simulations with 2D crystals may be achieved by making use of transverse motional modes to generate entanglement~\cite{Yoshimura2015,Richerme2016,Wang2015}. Towards this goal, recent experiments have demonstrated good isolation between transverse  motion~\cite{D'Onofrio2021,Qiao2021}, where micromotion may be minimized and ground state cooling has been achieved, and radial motion, where excess micromotion is present, confinement is weaker, and efficient cooling may be difficult.

\begin{figure*}[]
\captionsetup{width=\textwidth}
\includegraphics[width=\textwidth]{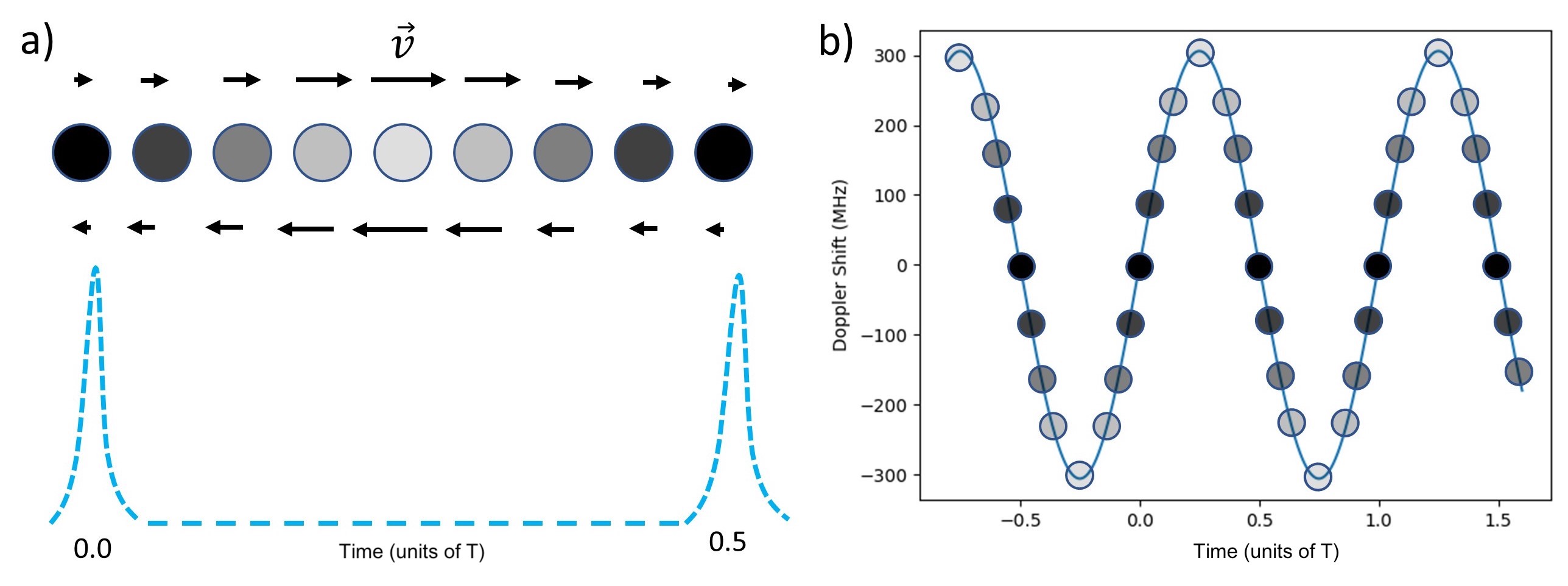}
\caption{Trapped ion excess micromotion effect on laser absorption. a) An ion oscillating back and forth at frequency $\Omega$ in the direction of the wave vector $\vec{k}$ of a cooling beam leads to a time dependent Doppler shift.  The transparency of each dot represents the magnitude of it's velocity $\vec{v}$, also indicated by the arrows. By pulsing a laser on (laser intensity indicated by the dashed line) during only a small portion of the ion's trajectory when the velocity is minimal (darkest points), one can selectively cool a smaller velocity class. Time is measured in units of the RF period $T$.
b) Doppler shift of a 493~nm light in the reference frame of a Ba$^+$ ion undergoing micromotion at 8~MHz with 3~$\mu$m amplitude. Instantaneous Doppler shifts exceed 300~MHz, which is large compared to the 15~MHz natural linewidth of the transition.}

\end{figure*}

To reach the low temperatures required for quantum information experiments, one must effectively cool along each trap axis. This necessitates having at least some component of the cooling beams point in the direction of excess micromotion, which may have a dramatic effect on the steady-state Doppler cooling \cite{Berkeland1998,Devoe1989}.To avoid these adverse effects one can cool largely along an axis where no micromotion is present \cite{Wang2020,Szymanski2012}, such as with lateral 2D crystals. However, at least some of the cooling beam's k-vector must point in a direction where micromotion is present, since otherwise thermal motion in all directions is not cooled. As the size of crystals scales up, this effect may become significant. Moreover, for radial 2D  crystals \cite{D'Onofrio2021,Xie2021,Kato2022}, and for 3D crystals, it is generally not possible to isolate a direction in space with no excess micromotion. Therefore, for large trapped ion crystal, it may be necessary to directly take into account and overcome the detrimental effects of micromotion on Doppler cooling.

 Due to  micromotion, the cooling laser frequency in each ion's rest frame is continuously Doppler-shifted, by varying amounts across a crystal. This causes the absorption spectrum and range of frequencies for which steady state cooling is efficient to  vary for different ions. It was suggested to use power-broadened and far-detuned cooling beam to allow more even cooling across a wide range of micromotion amplitudes \cite{Devoe1989}. In recent experiments, we implemented a two-tone Doppler cooling \cite{Kato2022} and were able to stabilize larger radial 2D crystals covering a broad range of micromotion. Multi-tone Doppler cooling may be an avenue to stabilizing even larger crystals. However, effective Doppler cooling of crystals where ions have both large and differing amplitudes of micromotion parallel to the cooling beam's k-vector remains an outstanding challenge.

In this paper, we propose using pulses of Doppler cooling light synchronized with the trap RF phase to cool ions undergoing micromotion, as illustrated in Fig. 1. There exist two points per RF period $T$ where the velocity $v=0$ for all ions simultaneously. We propose to use $\sim$ns laser pulses synchronized with the nodes in the micromotion velocity (dashed line in Figure 1.a) in order to narrow the range of ion speeds which need to be addressed by the cooling beam. We then show that this technique can be useful when cooling multiple ions with differing amounts of micromotion using a single-tone  laser beam. 

Pulsed Doppler cooling has been used before with the intention of broadband cooling~\cite{Blinov2006} or generating frequency comb
teeth deep into the ultraviolet (UV) range, offering to reduce
the complication associated with harmonic generation of light~\cite{Udem2016,Jayich2016}. However, laser pulses have not been used to cool trapped ions synchronized with the trap RF. In addition, micromotion-synchronized frequency modulation has been used previously to compensate for the effect of micromotion on Raman transitions in surface traps~\cite{Negnevitsky2018,Wan2019,Erickson2021}. The frequency modulation scheme may be effective in cooling trapped ions with the same amplitude and phase of micromotion, but breaks down for 2D crystals, where ions have different amplitudes of micromotion with a phase flip at the trap center.
\section{Pulsed Cooling }

Previous approaches to modelling Doppler cooling under micromotion have relied on time averaging steady state solutions to the Schrodinger's equation or the optical Bloch equations by sampling velocities over a period of micromotion to produce an atomic absorption spectrum \cite{Kato2022,Devoe1989,Berkeland1998}. Yet in the presence of significant micromotion, the cooling is not steady state since ions may be experiencing Doppler shifts much larger then the linewidth of the atomic transition $\Gamma$ and rapidly changing on a timescale similar to the excited state lifetime $\tau$. Moreover, pulsed lasers cause frequency combing effects and fast intensity changes that cannot be captured in the steady state. Therefore, to understand how pulsed Doppler cooling works, we numerically solve the time-dependent optical Bloch equations (see appendix for details).

\begin{figure*}
\includegraphics[width=0.95\textwidth]{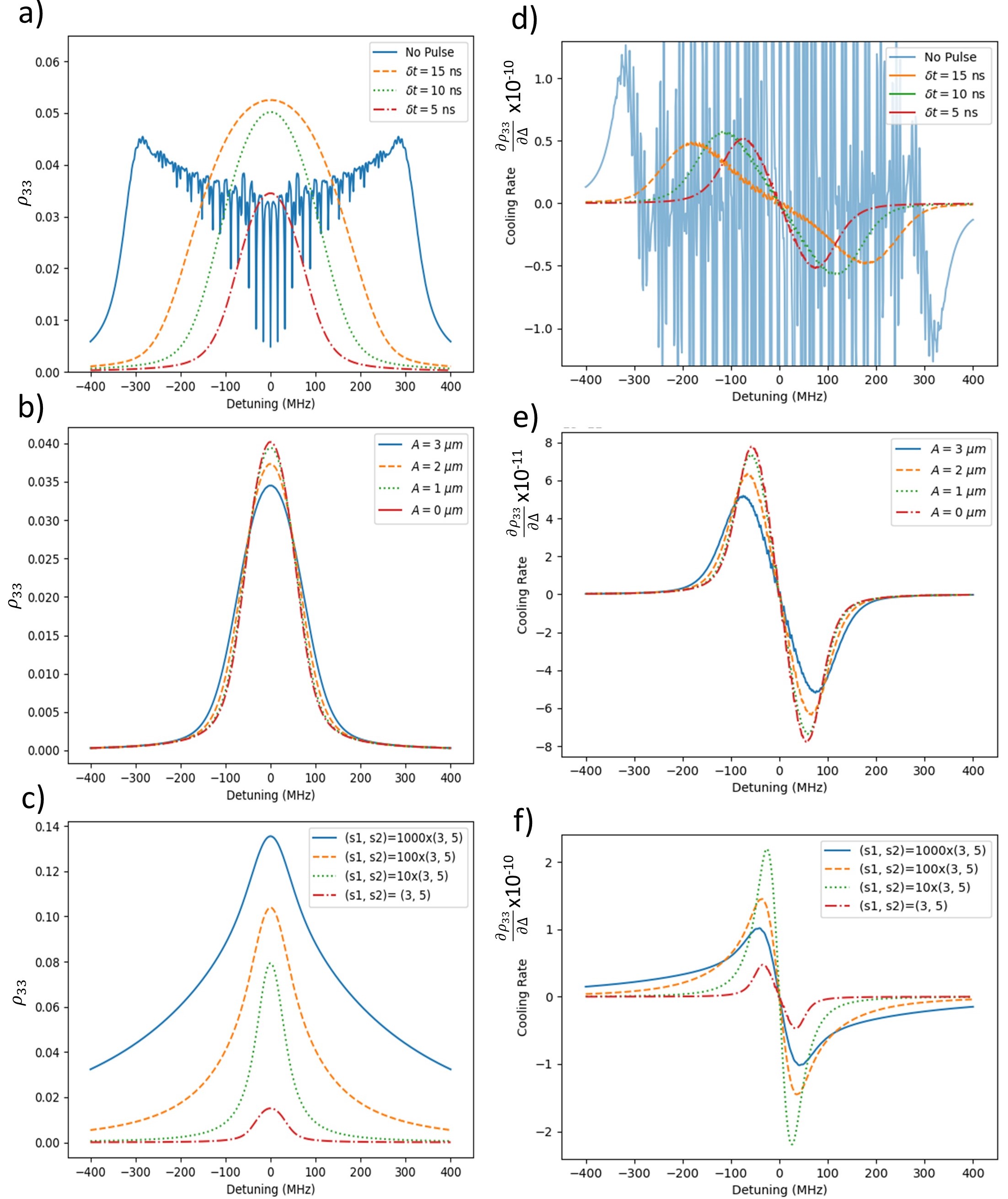}
\caption{\label{fig:Figure_2} (a-c) Benefits of using pulsed Doppler cooling. (d-f) Cooling rates corresponding to the scenarios in panels (a-c). The excited state population $\rho_{33}$ is plotted as a function of the cooling laser detuning $\Delta$ for: a) different pulse widths $\delta t$ and fixed amplitude of micromotion $A=3$ $\mu $m, b) fixed pulse width $\delta t=5$~ns and different micromotion amplitudes $A$. This is close to the condition of Eq. 3. c) fixed $\delta t=10$ ns, $A=0$ $\mu$m and different cooling laser intensities listed in terms of the saturation intensities s1 and s2 of the cooling and the repump lasers, respectively.
We note that narrowing the pulse width narrows the absorption spectrum considerably, leading to a more uniform spectrum across ions with differing amounts of micromotion. This may lead to an improved average cooling rate over the entire crystal. Saturation affects the cooling spectrum in a similar way to the steady-state case, but at higher average laser powers. 
}
\end{figure*}

Consider a crystal of $^{138}\text{Ba}^+$ ions undergoing micromotion at a frequency $\Omega=2 \pi \times 8$~MHz (period $T=2\pi/\Omega=125$~ns), interaction with the 493-nm Doppler-cooling laser (transition linewidth $\Gamma_{31}=2\pi\times 15$~MHz) and 650-nm repump laser (transition linewidth $\Gamma_{32}=2 \pi \times5$~MHz), see appendix, Fig. 5. These values are similar to what our experiment is capable of, and are well representative of the regime where $\Omega<\Gamma_{31}$. Each ion experiences oscillations around a fixed point described by $ \vec{r}=A \text{cos}(\Omega t)$ where $A={q r_s}/{2}$ is the micromotion amplitude. Here, $r_s$ is the displacement of the equilibrium position of the ion from the trap center, and $q$ is the relevant Matthieu parameter. Hence, the instantaneous velocity is
$v=A\Omega\text{sin}(\Omega t)$, and the instantaneous Doppler shift in the ion's rest frame is $\vec{k}\cdot \vec{v}=kA\Omega \text{sin} (\Omega t)$, where $\vec{k}$ is the laser wave vector. An example of this is shown in Figure 1.b for an ion with micromotion amplitude $A=3$~$\mu$m.

We model the behaviour of the excited state population, $\rho_{33}$, as shown in Fig. 2. In order to compute the absorption spectra, we time-average the solutions over many periods of micromotion. In Figure 2.a we show the benefits that can be obtained by pulsing the cooling beams, as opposed to continuous cooling, for different pulse widths. The absorption spectrum for continuous cooling obtained via this method (solid line in Fig. 2.a) is similar to those found using the steady-state model \cite{Kato2022}, yet the adverse effects due to coherent population trapping (CPT) are far more pronounced. As the amplitude of micromotion becomes large, multiple dips appear at multiples of $\Omega$. This is in contrast to the steady-state solutions, where these CPT features smooth out as the spectrum becomes power broadened \cite{Kato2022}. In Figure 2.b we plot the absorption spectra for the fixed laser pulse width of 5~ns and various micromotion amplitudes. We note that the absorption line width remains essentially unchanged as the micromotion amplitude increases from 0 to 3~$\mu$m. In Fig. 2.c we show the effect of power broadening in pulsed cooling, which shows the expected behavior of the line width increasing at higher laser intensities. The derivatives of the absorption curves in Fig. 2 (a-c), which are proportional to  the cooling rates, can be found in panels (d-f).

In order to cool the ions effectively we must take into consideration the level of velocity selection due to the pulse width, the frequency combing effects from the pulse train, the influence on CPT dips, and the saturation effects. Each has a substantial impact on cooling efficiency and must be considered individually and with respect to each other.

First, we consider the effects of CPT when cooling using pulsed lasers. For a 3-level $\lambda$-system such as $^{138}\text{Ba}^+$, the combined effect of the CPT and the frequency combing leads to a complicated absorption spectrum that does not yield good cooling. Instead, by alternating the cooling and the repump pulses at frequency $\Omega$ (as can be seen, for example, in Fig. 3.b), we can eliminate the CPT effect altogether. In this case the cooling transition and repump are never illuminated simultaneously and we have that $1/\Gamma=\tau<<T$ and the absorption spectrum resembles a 2-level system (for the following discussion $\Gamma=\Gamma_{31}$). The CPT dips no longer exist and do not effect the absorption spectrum under micromotion.  We also note that for a true 2-level system such as $^{114} \text{Cd}^+$, the CPT effects would not be a consideration.

\begin{figure*}[]
\includegraphics[width=\textwidth]{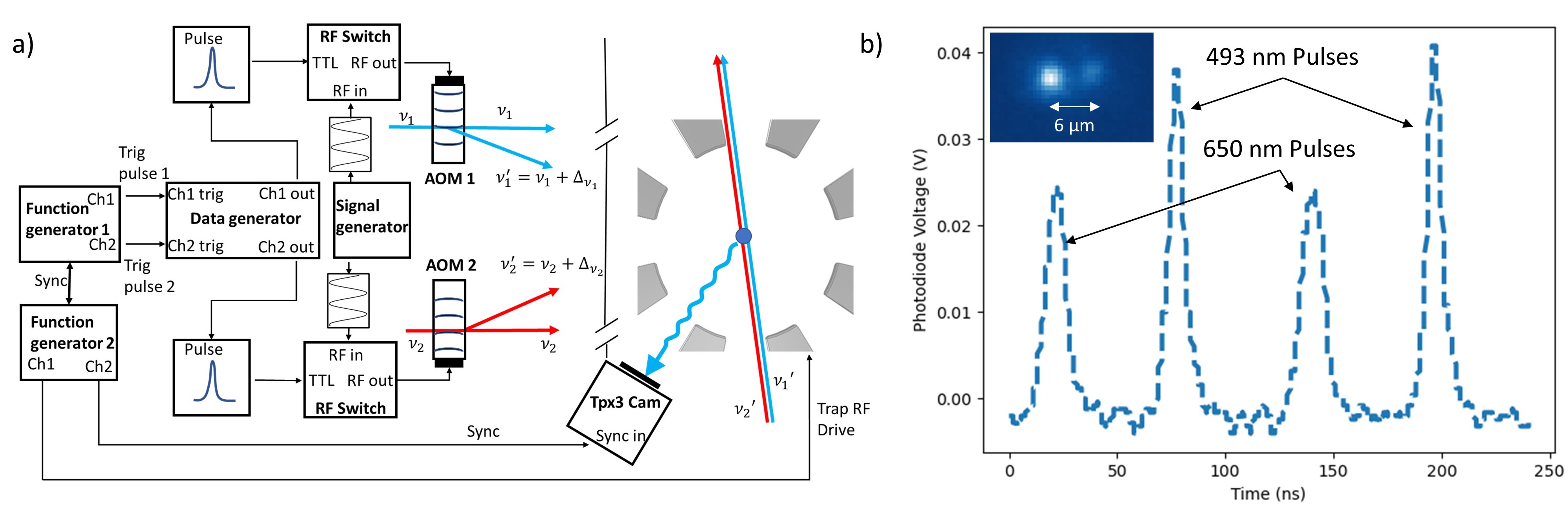}
\caption{\label{fig:Figure_3} Schematic for pulsed laser cooling and temporal profile of laser pulses. a) The schematics of the experimental setup. Optical pulses at frequency $\nu_{1}'$ and $\nu_{2}'$ are generated via AOMs, using the first diffracted order. The two pulsed beams are combined after the AOMs and sent to the trap. RF pulses that drive the AOMs are generated by sending short TTL pulses from a programmable data generator into fast RF switches. Two phase-locked RF synthesizers are used to drive the trap RF, to synchronize data acquisition and time measurement with the fast camera, Tpx3Cam, and to trigger the pulses from the data generator. b) Temporal profiles of the combined 493~nm and 650~nm optical pulses used in the experiment measured with a fast photodiode. The pulses are found to have $\delta t$ of 10 and 13~ns respectively for 493~nm and 650~nm light. Inset: single ion cooled using laser pulses synchronized with the trap RF. Using this technique, the ion is illuminated only at its endpoints of the micromotion trajectory. The micromotion amplitude is measured to be $\sim 3$~$\mu$m, or $\beta\sim39$. }
\end{figure*}

Next to consider are the frequency comb effects, which can also be detrimental to the cooling process. Consider a Gaussian laser pulse with a standard deviation $\sigma$, and thus a full width half maximum (FWHM) $\delta t=2.36\sigma$ (see appendix). Narrow pulses give the best velocity selection, but too narrow and combing effects become a nuisance, as more teeth fit into the comb envelope and broaden the absorption spectrum. The repetition rate $f_r=\Omega/2\pi$ with pulse width $\delta t$ produces a comb with teeth separated by $f_r$ and envelope width $\delta\nu=1/{\delta t}$. If $\Gamma>\Omega$ then the teeth are not resolved and the effect is a broadening of the absorption spectrum to width $\Gamma_{c}\sim \delta \nu$ , where $\Gamma_c$ is the approximate comb broadened linewidth. In our case, we restrict our attention to the case $\Gamma>\Omega$, since it is experimentally feasible given the range of trap strengths and the ion species used.

On its own, pulsing does not improve the Doppler cooling. However, in the presence of significant micromotion the pulsing can be effective in narrowing the width of the absorption spectrum.
The approximate absorption linewidth due to micromotion broadening $\Gamma_{\Omega}$ is twice the maximum Doppler shift $\vec{k}\cdot \vec{v}_{max}$. Yet in addition to this broadening, the absorption spectrum is also distorted. When we pulse the lasers such that $\Gamma_{c}<\Gamma_{\Omega}$, improvements can be seen. This  can be thought of as an effective reduction of the micromotion modulation index $\beta={k}v_{max}/\Omega$. For an ion cooled by a pulsed laser, velocity selection using a pulse with a width $\delta t$ leads to $v'_{max}=A\Omega^2\delta t$, while for a continuously cooled ion, $v_{max}=A\Omega$. Therefore we obtain an expression for a reduced modulation index $\beta'$

\begin{equation}
\frac{\beta'}{\beta}=\frac{v'_{max} }{ v_{max}}=\Omega \delta t
\end{equation}
in the limit $\delta t/T<<1$.
Thus, we find the approximate condition where pulsing is effective when $\Gamma_{c}\leq\Gamma_{\Omega}$, that is 

\begin{equation}
2 \pi\delta \nu \leq2 \beta ' \Omega
\end{equation}
leading to the condition 
\begin{equation}
\delta t\geq\sqrt{\frac{\pi}{ \beta \Omega^2}}.
\end{equation}
For example, with $A=3$~$\mu m$ at $\Omega=8$~MHz, we have that $\delta t\geq5.6$~ns. In this limit, the cooling rate of a single ion does not improve, but the average cooling rate across all ions in a crystal can be improved dramatically when considering a single cooling beam.

Finally, one must consider the power broadening. Since the laser is pulsed, only a fraction of the power is delivered to the ion in contrast to the continuous wave laser cooling. Saturation and saturation intensity are steady-state concepts, yet some intuition can be gained by replacing the saturation parameter $s$ with a reduced saturation parameter, $s'= s (\delta t /T)$. The power broadened linewidth is then reduced to $\Gamma_{s'}=\Gamma\sqrt{1+s'}$. 

We note that the scenario would be quite different in the case $\gamma < \Omega$. In this case the effect of micromotion is to cause resolved sidebands in the absorption spectrum spaced at $\Omega$~\cite{Devoe1989,Berkeland1998}. Therefore, pulsing the laser at $\Omega$ would add comb teeth spaced equally to the micromotion sidebands. One could align teeth with sidebands in order to maintain cooling. This would not require selecting nodes in the velocity due to micromotion. Moreover, since the comb teeth are resolved, the features of the absorption spectrum may not be broadened. We leave this regime to be further explored.

In summary, shorter pulses will help to cool an ion until the condition in Eq. 3 is reached. Therefore, when considering a large crystal, one should consider the distribution of amplitudes of micromotion. Shorter pulses will help cool ions where the amplitude of micromotion is greater and where a large range of micromotion-induced Doppler shifts needs to be covered, but it will not help if the micromotion amplitude is small, or if cooling to near the Doppler limit is needed. However, average cooling rates across all ions may be improved, leading to an overall lower crystal temperature.

\section{Experiment and methods}

The ion trap and the Doppler cooling setup is similar to the one described in \cite{Kato2022,Ivory2020}. We cool  $ {}^{138} \text{Ba}^+$ using a beam at 493~nm and repump transition at 650~nm to address the $6 S_{1/2}\leftrightarrow 6P_{1/2}$ and  $5 D_{3/2}\leftrightarrow 6P_{1/2}$ transitions, respectively. A small magnetic field is in the transverse direction defines the quantization axis at approximately
$90\degree$ with respect to the polarization of the cooling beams, a configuration that avoids optical pumping into a
metastable dark state~\cite{Berkeland2002,Oberst1999}.
Pulses of cooling light are created using Crystal Technology 3200 series Acousto-optic modulators (AOM's) with a central frequency of 200~MHz, driven by a HP 8657D signal generator. The RF signal at 200~MHZ is sent to the AOMs through Minicircuits ZASW-2-50DR+ switches that have a typical rise/fall time of 6~ns. The switches are controlled using a DG2020A Data Generator that has a rise/fall time of 2~ns.

The trap RF at $\sim$8~MHz is produced by a Siglent SDT 2042X arbitrary waveform generator. The second channel of the generator is used to send trigger pulses at 100~kHz to the TimePix3 camera to synchronize the camera internal timing with the trap RF. A second Siglent SDT 2042X arbitrary waveform generator is phase-locked to the first one using the 10~MHz time base, and is used to trigger the DG2020A Data Generator. Two channels of the second Siglent SDT 2042X with adjustable phase allow the relative timing of the 493~nm and 650~nm laser pulses to be tuned, as well as the relative phase of both pulses with respect to the trap RF. A schematic of the setup is shown in Fig. 3.a.

Laser pulses are measured using a Thorlabs PDA10A fast photodiode and Rigol DS 1202-ZE oscilloscope. The pulse parameters are roughly adjusted so that the pulses are separated to as close to $T/2$ as possible given the granularity of DG2020A ($\sim 5$~ns). The width of the pulses is determined to be 10~ns and 13~ns for the 493~nm and the 650~nm lasers, as depicted in Figure 3.b. For each beam, the total power is measured to be approximately 1/10th of the continuous wave power.

While initially trapping, ion crystal formations are observed using an electron multiplying charge-coupled device (EMCCD) camera as described in \cite{Ivory2020}. For fluorescence data collection, the ion formations are imaged onto the intensified complementary metal-oxide-semiconductor (CMOS) camera, Tpx3Cam, using the same setup as in \cite{Kato2022}, providing single-photon sensitivity and time-stamping functionality in each pixel with precision of 1.6~ns~\cite{Fisher2016,Nomerotski2019}. A time-to-digital-converter (TDC) with  260~ps time resolution built into the camera receives and time-stamps pulses that are synchronized with $\Omega$ in order to observe micromotion, as described in \cite{Zhukas2021}.

\begin{figure*}[!t]
\includegraphics[width=\textwidth]{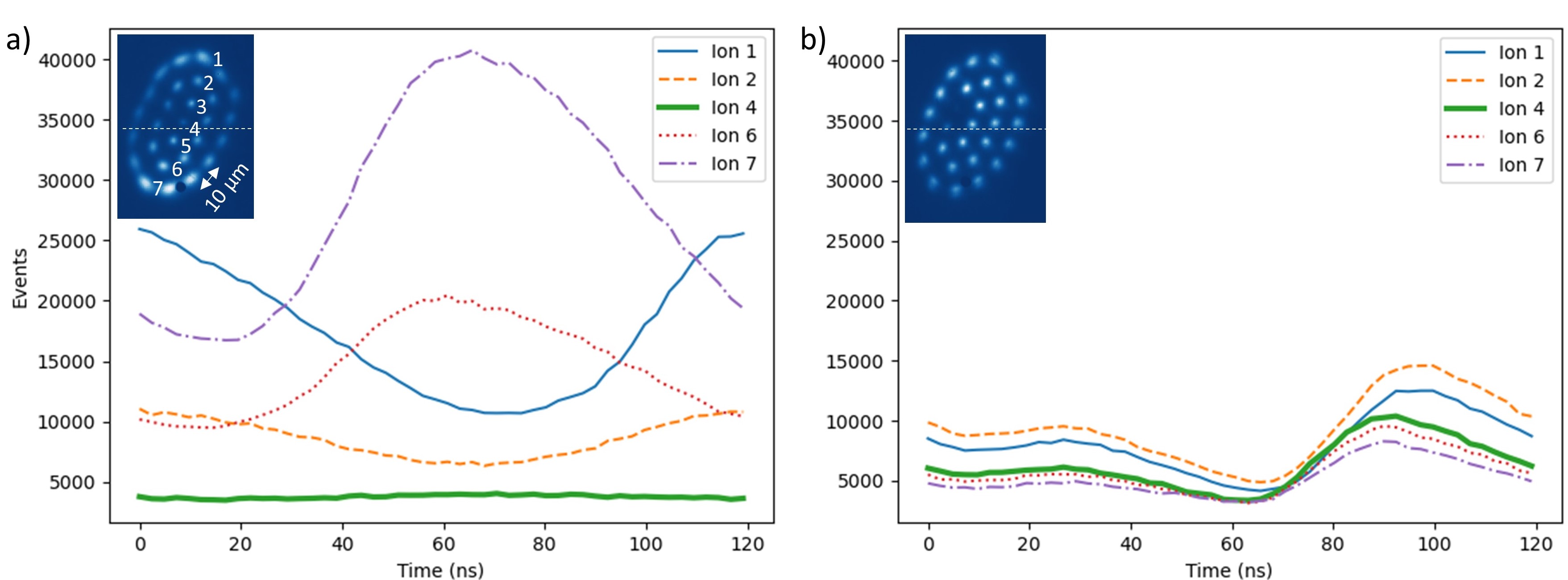}
\caption{\label{fig:Figure_4}  Scattering rates of single ions over the RF period $T$ for the continuously cooled (a) and pulsed cooled (b) 28-ion crystal centered near the trap RF null, taken with the TimePix3 camera. The trap frequencies are $(\omega_x,\omega_y,\omega_z)=2 \pi \times (193,266,737) $~kHz. Insets are integrated EMCCD images of the crystals. The laser propagates in the $Y$ (vertical) direction in these images, and the dashed lines represent $Y=0$. Ion spacing is approximately 10~$\mu$m. Each curve represents fluorescence of a single ion, as labelled in the inset of (a). These ions are chosen to represent the wide distribution of micromotion amplitudes across the crystal. Vertical scales in arbitrary units are identical in both panels. a) Relative scattering rates of ions 1,2,4,6 an 7 over an RF period using continuous laser cooling. Broad variations of the scattering rates due to Doppler shifts can be seen between different ions with different micromotion amplitudes. b) Same as (a), but with pulsed laser cooling. Scattering rates for different ions are now nearly identical. The scattering rate variation over the trap period is due to the pulsed nature of the lasers cooling. For the central ion (ion 4) the average fluorescence rate actually increases by a small amount in the case of pulsed cooling, since the laser frequency can be brought up much closer to resonance.}
\end{figure*}

We find that it is necessary to use full power (i.e. non-pulsing) cooling laser beams when first forming the ion crystals with more than a few ions. We cool crystals using a red-detuned lasers as used in previous experiments \cite{Ivory2020,Kato2022}. Once the crystal has formed, the pulsing is turned on. 

As soon as the pulsing has begun, if the timing of the pulses is correct, the ions become quite dark. This is because  the cooling laser frequency is significantly (a few hundred MHz) red-detunned to enable continuous Doppler cooling. The main cooling beam at 493~nm must be quickly tuned up towards the resonance by ~100 MHz. If the crystal appears bright in some areas but not others and easily melts, the phase of the pulses relative to the trap RF is incorrect causing only certain ions with the correct velocity range to fluoresce brightly. Typically, these crystals do not last long before melting. If the relative phase is nearly correct, the crystal appears mostly uniform in brightness and does not melt, and the cooling beams are successfully illuminating ions only near $v=0$. 

To fine-tune the relative phase of the laser pulses to the trap RF, we trap and a single ion and displace it from the trap center by applying a bias voltage to one of the trap's sector electrodes to induce a large amount of micromotion for the ion. We then dynamically adjust the delay between the laser pulses and the trap RF zero-crossing while monitoring the ion image with the camera. The correct phase is achieved when the ion image appears as two spots corresponding to the ion positions at the micromotion extrema, as shown in the Fig. 3.b inset. The frequencies of the cooling and repump lasers are then tuned up as far as possible towards the resonance without melting the crystal. When correctly configured, the pulses select velocities near $v=0$ and the lasers can be tuned farthest into the blue.

We trap and cool a 28-ion, 3-shell 2D crystal centered near the trap's RF null with and without the pulsed laser method, as shown in Fig. 4. The cooling laser direction is vertical in the ion crystal images, and the location $Y=0$ of the trap's RF null in that direction is indicated by the dashed white line. Ions in this crystal experience micromotion up to $A\sim1.5$~$\mu m$ ($\beta=19$), with about half the ions above the $Y=0$ line oscillating out of phase with the ions below. We observe that all ions have a relatively uniform scattering rate (and therefore brightness) when using the pulsed laser method as compared to the continuous laser cooling, where a broad variations of scattering rates are observed. We also note that scattering rates of ions above and below the $Y=0$ line (e.g. ions 1 and 7 in Fig. 4.a) are out of phase for the continuous laser cooling, as expected. For the pulsed laser cooled ions, this phase flip in fluorescence is not observed, since the scattering rate is now determined by the pulse profiles. In addition, the crystal temperature is reduced modestly from approximately 15~mK in the continuous laser cooling to 13~mK in the pulsed laser cooling case, as determined from the spatial extent of the ions using molecular dynamics (MD) simulations~\cite{Kato2022}.

To fully explore the benefits of pulsed Doppler cooling, we need to reduce the pulse widths down to the limit imposed in Eq. 3. Since we are limited to $\delta t \sim 10$~ns by the available hardware, the cooling rates and the ion fluorescence profiles are still not fully uniform across crystals formed in this trap. Faster RF switches and AOMs, and lower trap RF drive frequencies may enable this region of pulsed laser cooling to be more fully explored. Given the range of micromotion that can be addressed using this technique, it may be possible to perform Doppler cooling of $n>100$ ions.

\section{Conclusion}
We have proposed and demonstrated the use of pulsed lasers to perform velocity-selective Doppler cooling of ions undergoing micromotion. The pulsed cooling allows ions with differing amounts of micromotion to be addressed by a single cooling beam, and improves the average cooling rate even when the amplitude of micromotion becomes significant. Ultimately, we are limited by available laser power and minimum pulse duration. In future experiments, we intend to use shorter laser pulses to explore the limitations of pulsed Doppler cooling.

\section{Acknowledgements}

The authors would like to thank Vasileos Niaouris for help with the data generator.  A.K. acknowledges funding from the  University of Washington Department of Physics General Excellence Award. A.N. acknowledges support from the U.S. Department of Energy QuantISED award and BNL LDRD grant 19-30.

\bibliography{refs}

\section{Supplemental Material}

\subsection{Optical Bloch Equations}
Consider an ion undergoing micromotion at frequency $\Omega$, illuminated by lasers driving transition a cooling transition $\ket{1}\leftrightarrow\ket{3}$ with natural linewidth $\Gamma_{31}$ and a repump transition$\ket{2}\leftrightarrow\ket{3}$ with natural linewidth $\Gamma_{32}$ (Fig. 5). 

\begin{figure}[b]
\includegraphics[width=0.5\textwidth]{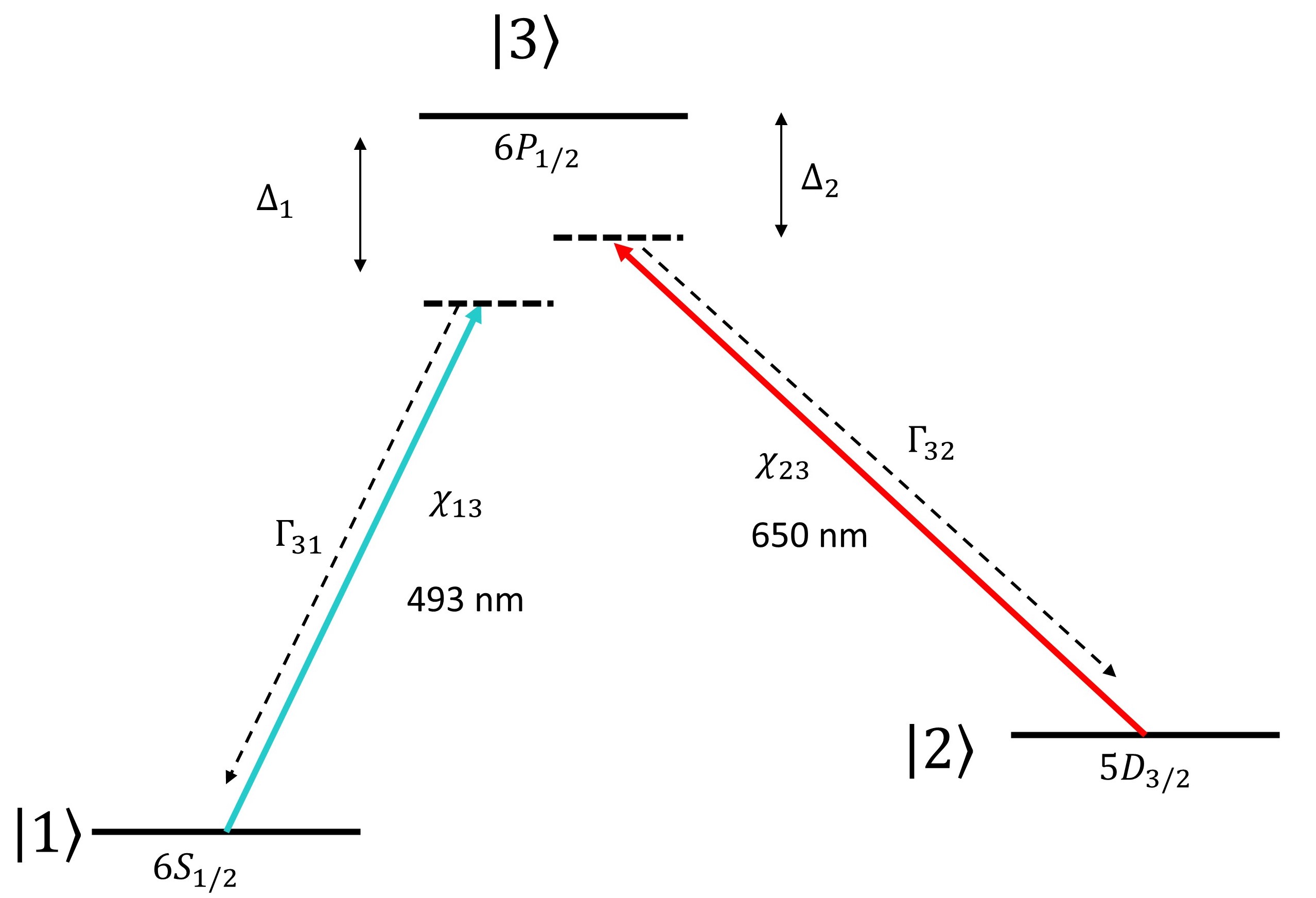}
\caption{\label{fig:Figure_5} $^{138}\text{Ba}^+$ energy levels, transition linewidths and applied laser fields.}
\end{figure}

We use the Linblad master equation to derive the optical Bloch equations for this system in order to evaluate the evolution of the density matrix $\rho$
\begin{equation}
   \dot{\rho}=\frac{i}{\hbar} \left[ H,\rho\right]+\mathcal{L}(\rho),
\end{equation}

where H is the Hamiltonian, and $\mathcal{L}$ is the Liouvillian defined by 

\begin{equation}
   \mathcal{L}(\rho)= \sum_{i}^{} \frac{1}{2}(2C_n \rho C_n^{\dagger}-\rho C_n^{\dagger}C_n-C_n^{\dagger}C_n\rho).
\end{equation}

We ignore the dephasing and instead only consider radiative damping, which is the only loss terms that come from $\Gamma_{31}$ and $\Gamma_{32}$. The collapse operators $C_n$ are $\sqrt{\Gamma_{31}}\ket{3}\bra{1}$ and $\sqrt{\Gamma_{32}}\ket{3}\bra{2}$. The Opical Bloch Equations are

\begin{equation} 
    \dot{\rho_{11}}=i\frac{\chi_{13}(t)}{2}(\rho_{31}-\rho_{13})+\Gamma_{31}\rho_{33} \\
\end{equation}
\begin{equation}
\dot{\rho_{22}}= i\frac{\chi_{23}(t)}{2}(\rho_{23}-\rho_{32})+\Gamma_{32}\rho_{33}\\
\end{equation}
\begin{equation}
    \dot{\rho_{33}} = -\dot{\rho_{11}}-\dot{\rho_{22}} \\
\end{equation}
\begin{equation}
    \dot{\rho_{21}} = i\left((\Delta_{1}(t)-\Delta_2(t))\rho_{21}+\frac{\chi_{23}(t)}{2}\rho_{31}-\frac{\chi_1}{2}\rho_{31}\right) \\
\end{equation}
\small

\begin{equation}
\dot{\rho_{31}} = i\left(\frac{\chi_{23}(t)}{2}\rho_{21}+\Delta_1\rho_{31}+\frac{\chi_{13}(t)}{2}(\rho_{11}-\rho_{33})-\frac{1}{2}(\Gamma_{31}+\Gamma_{32})\rho_{31}\right)
\end{equation}

\begin{equation}
    \dot{\rho_{32}} =  i\left(\frac{\chi_{13}(t)}{2}\rho_{12}+\Delta_2\rho_{32}+\frac{\chi_{23}(t)}{2}(\rho_{22}-\rho_{33})-\frac{1}{2}(\Gamma_{31}+\Gamma_{32})\rho_{32}\right).
\end{equation}

\normalsize
Here, $\chi_{ij}(t)$ are the Rabi frequencies of the transitions indicated in Fig. 5, and $\Delta_{1,2}$ are the detunings of the 493~nm and  650~nm lasers from resonance, respectively.

 The equations are solved using a 4th order Runge-Kutta algorithm, keeping time steps small compared to $T$ (i.e. much smaller than 1~ns). We let the system evolve until it reaches a form that is periodic over the RF period, typically about 50$T$. We then let the system evolve for additional 100 $T$ to reach an equilibrium state.

To introduce the effect of a pulsed laser, we modulate the intensity of the 493~nm and  650~nm lasers, and therefore $\chi_{ij}(t)$. We consider pulses with a Gaussian temporal profile, such that the intensity of the pulse $I$ is given by 

\begin{equation}
I(t)= I_{peak}e^{-\frac{(t-t_p)^2}{2 \sigma^2}},
\end{equation}
where $I_{peak}$ is the maximum intensity. The pulses are centered at the desired time $t_p$, with $\sigma$ being the standard deviation of the Gaussian pulse (thus, the FWHM is $\delta_t = 2.36 \sigma$). The ion's velocity crosses zero at times $t_n=nT/2$ where n is an integer. Therefore, the lasers can be pulsed at at the rate $\Omega$ or $ 2\Omega$ while selecting velocities near $v=0$. For reasons described in the main text, we choose to pulse each laser at the rate $\Omega$, with a phase shift of $\pi$ between the two beams.

\end{document}